# The Thickness of Current Sheets and Implications for Coronal Heating

James A. Klimchuk
James E. Leake
Lars K.S. Daldorff [*]
Craig D. Johnston [**]

Heliophysics Science Division, NASA Goddard Space Flight Center, Greenbelt, MD, USA

[*] Also Catholic University of America, Washington, DC, USA
[**] Also George Mason University, Fairfax, VA, USA

The thickness of current sheets is extremely important, especially as it relates to the onset of fast magnetic reconnection. Onset determines how much magnetic free energy can build up in a field before it is explosively released. This has implications for many phenomena on the Sun and throughout the universe, including the heating of the solar corona. Significant effort has been devoted to the question of whether equilibrium current sheets in realistic geometries have finite or zero thickness. Using a simple force balance analysis, we show why current sheets without a guide field (2D) and with a guide field that is invariant in the guide field direction (2.5D) cannot be in equilibrium if they have both finite thickness and finite length. We then estimate the conditions under which the tension of a curved line-tied guide field can facilitate equilibrium in 3D sheets that are finite in all dimensions. Finally, we argue that some quasi-statically evolving current sheets undergoing slow stressing – e.g., when the coronal magnetic field is subjected to photospheric boundary driving – may reach a critical shear, at which point they lose equilibrium, spontaneously collapse, and reconnect. The critical shear is generally consistent with the heating requirements of solar active regions.

## INTRODUCTION

Many explosive phenomena occurring on the Sun, within the heliosphere, and throughout the universe involve the slow buildup and sudden release of magnetic energy. Solar examples include flares (Kazachenko et al. 2022), coronal mass ejections (Chen 2011), jets (Raouafi et al. 2016), and the nanoflares that heat the corona to temperatures of several million degrees (Klimchuk 2015). In a typical scenario, slow forcing at the boundary of the system causes magnetic stresses to grow. Current sheets become thinner until eventually reaching a critical thickness whereupon fast magnetic reconnect sets in and energy is explosively released. On the Sun, the boundary forcing is provided by photospheric flows that displace the footpoints of coronal magnetic field lines. Chaotic flows associated with turbulent convection are especially relevant to coronal heating.

Half a century ago, Parker (1972) proposed that infinitely thin current sheets – also called singular current sheets and tangential discontinuities – must develop whenever continuous 3D magnetic fields (without separatrices or X-points) are subjected to continuous motions at a line-tied boundary. This provided a straightforward explanation for coronal heating because ubiquitous current sheets would be expected to form and reconnect. Parker called this process "topological dissipation."

Parker's picture is in fact problematic. Minimal stress is built up in the field – and minimal energy is released – if reconnection happens too readily, as would be the case if current sheets were singular when they first form. Even a high occurrence frequency of weak events is inadequate to heat the corona because the time-averaged Poynting flux of energy pumped into the field by photospheric driving depends on the level of stress that is present (Klimchuk 2015).

The question of whether current sheets have zero or finite thickness is still being actively debated. Most authors disagree with Parker and conclude that the current sheets in line-tied 3D fields are not singular. The scenarios that have been investigated include quasi-static sheet formation from boundary driving (van Ballegooijen 1985; Antiochos 1987; Zweibel & Li 1987; Mikic, Schnack, & Van Hoven 1989; Craig & Sneyd 2005; Aulanier, Pariat, & Demoulin 2005; Zhou et al. 2018; Huang et al. 2022), dynamic sheet formation from instabilities (Longcope & Strauss 1994; Baty 1997; Huang, Bhattacharjee, & Zweibel 2010), and dynamic sheet formation from the relaxation of braided out-of-equilibrium fields (Wilmot-Smith, Hornig, & Pontin 2009; Pontin & Hornig 2015). Low (1992) and Ng and Bhattacharjee (1998) argue in favor of Parker. However, none of these studies can be considered definitive or universal.

Here we address the problem from a different approach. We examine the balance of forces within current sheets of finite thickness to determine whether and when equilibrium is possible. We first consider simplified 2D and 2.5D geometries before moving on to fully 3D sheets with line-tied boundary conditions, as applies to the corona.

While we emphasize coronal heating, our results are general. We note that reconnection occurs at different current sheet thicknesses in different physical environments. In nanoflares, the tearing instability that initiates reconnection is fast for sheets much thicker than kinetic scales – so a resistive MHD approach is valid – while in the magnetosphere, kinetic effects are fundamentally important at reconnection onset.

## 2D AND 2.5D CURRENT SHEETS

There exist many published two-dimensional solutions for equilibrium current sheets without a guide field (2D) and with a guide field that is invariant in the guide field direction (2.5D).[1] These sheets are either infinitely long (Harris 1962) or infinitely thin (Green 1965; Syrovatskii 1971; Priest 1985). Here, length and thickness refer respectively to the dimensions along and across the sheet in the plane of reconnection. We searched the literature for solutions that are finite in both length and thickness but were unsuccessful. This led us to wonder whether such solutions are possible, and, for the reasons given below, we have concluded that they are not.

The top panel of Figure 1 shows a 2D equilibrium current sheet of finite length and zero thickness (Green 1965). The sheet is the horizontal line at the center that is bounded by two Y-points. The field is potential everywhere except at the sheet itself and is oppositely directed above and below the sheet. Since there are no plasma forces, the Lorentz force is everywhere zero:

$$\frac{1}{c}(\bm{J} \times \bm{B}) = 0 \ . \tag{1}$$

The Lorentz force can be separated into two terms – one associated with magnetic pressure and the other associated with magnetic tension. These two forces exactly balance at all locations:

$$\nabla\left(\frac{B^2}{8\pi}\right) = \frac{1}{4\pi}(\bm{B} \cdot \nabla)\bm{B} \ . \tag{2}$$

The bottom panel of Figure 1 shows the magnetic pressure between two flux surfaces centered on the sheet. Gradients in pressure are offset by magnetic tension. Note the horizontal gradient in pressure along the sheet, with the highest pressure occurring in the middle ($x = 0$).

Magnetic tension is often associated with curved fields, but it also plays an important role in diverging and converging fields. To understand magnetic tension, it is instructive to consider the Maxwell stress tensor:

$$T_{ij} = \frac{1}{4\pi}\left(B_i B_j - \frac{1}{2}B^2 \delta_{ij}\right) \ . \tag{3}$$

The Lorentz force is equal to the divergence of the stress tensor, and the integral of the Lorentz force over a volume is, via the divergence theorem, identically equal to the integral over the bounding surface of the normal component of the stress tensor. The concept of magnetic tension refers to the idea that field lines tug on any surface through which they pass. The tugging force is directed along the field line. The first term in Equation 3 is associated with tension, and the second term is associated with magnetic pressure. Pressure forces act only perpendicular to a surface, whereas tension forces can have both perpendicular and transverse components.

---

[1] The guide-field direction is perpendicular to what is commonly referred to as the plane of reconnection, which contains the magnetic field components that participate directly in reconnection.

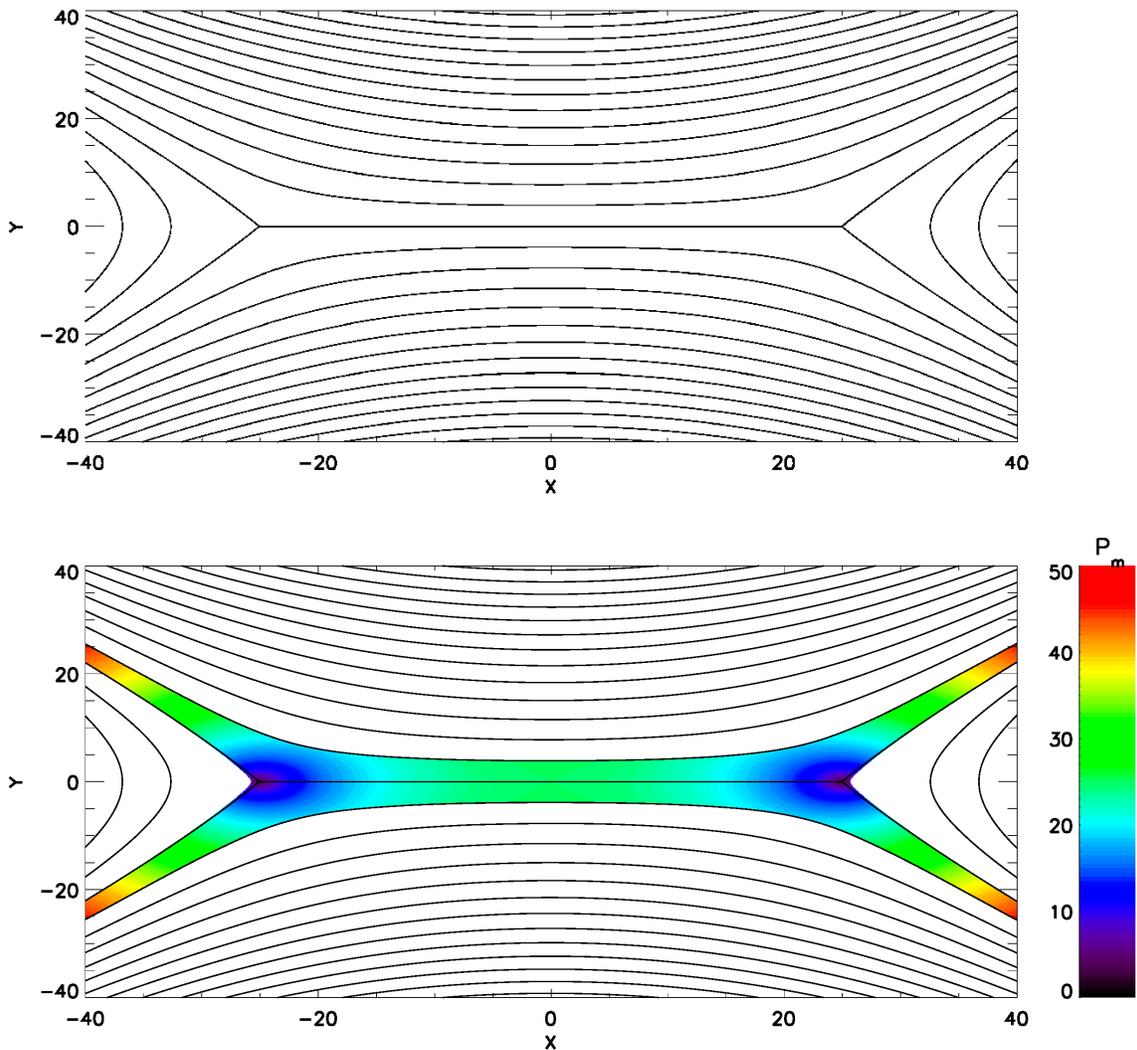

**FIGURE 1** Field lines of a 2D equilibrium magnetic field with an infinitely thin current sheet of finite length (top). Magnetic pressure between two flux surfaces equally spaced above and below the current sheet (bottom).

Consider a 2D curved field passing through a cube, as sketched at the top of Figure 2. Magnetic pressure pushes inward on all the faces of the cube. Tension pulls downward and to the left on the left face and downward and to the right on the right face. The leftward and rightward forces cancel, and the net tension force is downward. This may or may not be offset by a difference in pressure and/or vertical tension at the top and bottom faces. Examples of curved fields where the tension and pressure gradient forces are balanced can be seen near the Y-points in Figure 1.

The bottom of Figure 2 shows a diverging 2D field. The tension in the field lines is predominantly horizontal. However, moving from left to right, the horizontal component of tension decreases while the vertical component increases. Integrated over the faces of the cube, the vertical forces cancel, and there is a net tension force to the left. This is offset by a greater pressure pushing on the left face than the right face. This is the type of magnetic force balance that characterizes the colored region away from the Y-points in Figure 1.

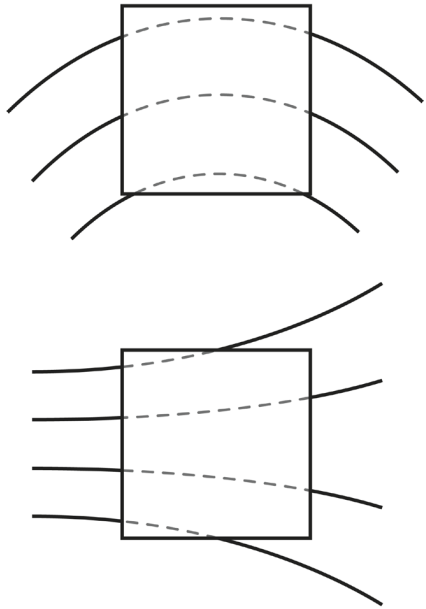

**FIGURE 2**  Cube threaded by a 2D curved magnetic field (top) and 2D diverging field (bottom).

Imagine that we create a finite thickness current sheet by replacing the magnetic field in the colored region with plasma having a gas pressure distribution equal to that of the removed magnetic pressure. The system is no longer in force balance. The horizonal pressure gradient remains, but there is no magnetic tension to offset it. The plasma will respond by flowing horizontally away from the middle of the sheet toward the Y-points in both directions. The sheet will become thinner as plasma evacuates and the external magnetic pressure squeezes in from above and below.

Can our hypothetical current sheet evolve to establish an eventual equilibrium with finite thickness? Because there is no magnetic field within the sheet, there is also no Lorentz force, so the plasma pressure must be uniform throughout the entire sheet. Force balance across the sheet boundary then requires uniform magnetic pressure just outside the boundary to match the uniform plasma pressure just inside. This means the field strength must be constant along the boundary.[2]

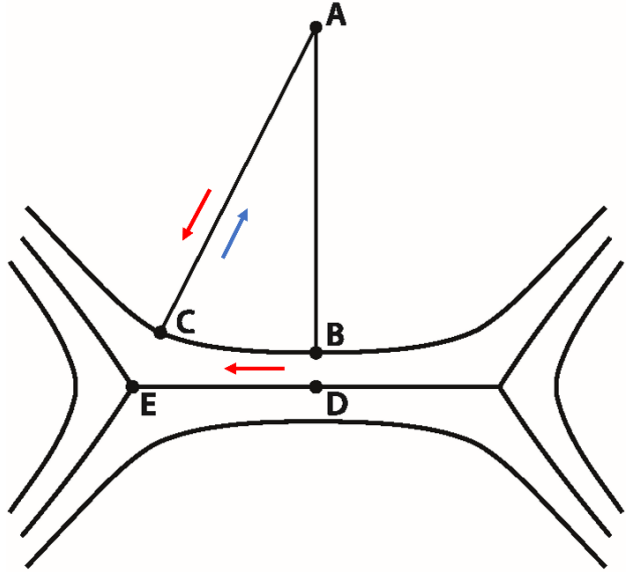

**FIGURE 3** Current sheet of finite thickness with points B and C on the sheet boundary and points D and E at the center of the sheet. Point A is vertically above B and D. Red and blue arrows indicate the directions of pressure gradient and tension forces, respectively.

Figure 3 shows a schematic sketch of the sheet. Points B and C are on the sheet boundary and point A is vertically above B. If the field strengths and magnetic pressures are the same at B and C, then the average magnetic pressure gradient along the path from A to B is greater than along the path from A to C, because the distance is shorter. To have equilibrium, the tension force must also be stronger along AB compared to AC. We can express the tension force (right-hand side of Equation 2) as $B^2/(4\pi R_c)$, where $R_c$ is the radius of curvature of the field. The radius of curvature is larger along AB than along AC and therefore the tension force is weaker, not stronger. Thus, there can be no force balance external to the sheet if the field strength is uniform along the sheet boundary. Force balance is possible only if the field strength decreases toward the Y-points: $B_C < B_B$. This is incompatible with uniform plasma pressure inside the boundary. Since no finite thickness equilibrium exists, the sheet must collapse to a singularity.

Note that the presence of external plasma would not change the situation. Magnetic pressure can be replaced by total pressure, and the argument above still holds. Force balance external to the sheet requires nonuniform total pressure just outside the boundary, which is incompatible with uniform plasma (total) pressure just inside. It should also be noted that plasma pressure is constant along every field line in an equilibrium because there is no Lorentz force parallel to the field. Thus, the plasma pressure is uniform just outside the boundary and the required nonuniformity is provided by the magnetic pressure. Finally, $\beta = 8\pi P/B^2$ is of order 1% in solar active regions, so the plasma has minimal impact on force balance in general. For these reasons, we do not include an external plasma in any of our analysis.

---

[2] A common misconception is that tension exerts a force perpendicular to the boundary if the boundary is curved. If the normal to the boundary has index i, then the tension term in the Maxwell stress tensor (Equation 3) – the first term – vanishes in all directions j because $B_i = 0$.

The hypothetical current sheet we have created by replacing all the magnetic field in the colored region of Figure 1 is illustrative but unrealistic. It should more properly be called a plasma sheet because the current is concentrated entirely at the sheet boundary, where the field ends abruptly. A more realistic situation is where the field and magnetic pressure decrease gradually to zero from the boundary toward the center ($y = 0$), while the plasma pressure increases gradually to a maximum at the center. The current is then smoothly distributed throughout the sheet. The well-known Harris sheet (1962) is of this type.

Figure 4 shows simple representations of the middle sections of four different current sheets, corresponding to the vicinity of $x = 0$ in Figure 1. Case I is the original field with an infinitely thin sheet. Case II is the plasma sheet discussed above. Case III is a current sheet where the field gradually transitions to plasma. Case IV is similar to Case III except that the in-plane magnetic field is replaced by a guide field component out of the plane, rather than by plasma. The magnetic field vector rotates smoothly by 180º across the sheet in Case IV. A modified version includes an additional uniform guide field, $B_{g0}$, and the field vector rotates by less than 180º. This is the situation for the current sheets associated with coronal heating, where a rotation of roughly 20º is needed to explain the energy requirements of active regions (e.g., Klimchuk 2015). Note that the sketches are only schematic and do not include the small horizontal gradients that would be expected and that are seen in Figure 1.

If the system has no variation out of the plane, then any guide field that is present must be straight. It therefore behaves like a plasma – it has pressure but exerts no tension force. Case III and Case IV are therefore effectively equivalent in 2D. We discuss the 2D force balance of Case III below, but plasma pressure and guide field pressure are interchangeable, so the conclusions apply equally to Case IV.

We start by considering the forces along the line between points E and C in Figure 3. E is at the Y-point and C is at the sheet boundary. The field is curved away from E and therefore a tension force is directed away from E. To balance this tension force, the total pressure must be smaller at E than C. In Case III, pressure is provided entirely by plasma at E and entirely by magnetic field at C, so $P_E < B_C^2/8\pi$, where $P$ indicates gas pressure. Force balance external to the sheet requires $B_C^2/8\pi < B_B^2/8\pi$, as discussed above. Vertical tension is very weak between B and D, so the total pressure must be similar at the two locations, implying $P_D \approx B_B^2/8\pi$. Following the chain, we find that $P_E < P_D$. However, because there is no Lorentz force along the magnetic field to balance this plasma pressure difference, equilibrium is not possible. Plasma will flow horizontally outward from the middle of the sheet, and the sheet will collapse to a singularity, just as in Case II. The same is true of Case IV.

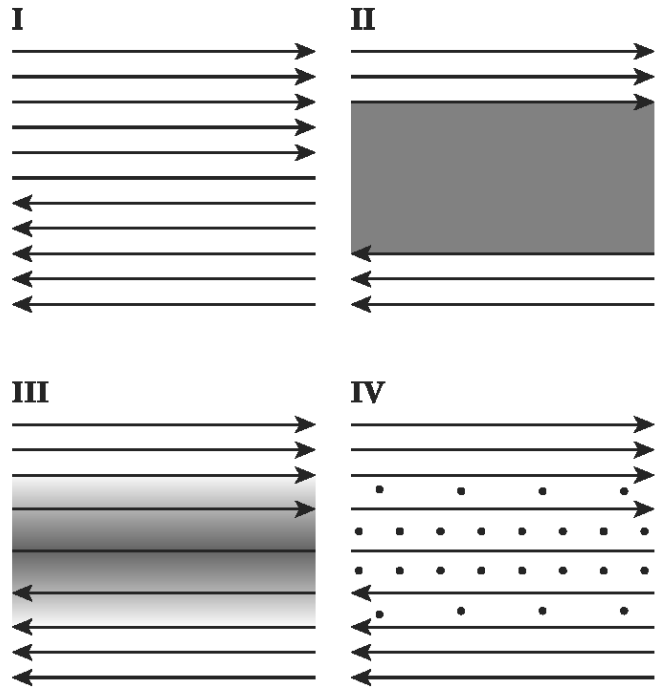

**FIGURE 4** Schematic representations of the middle sections of four current sheets: Case I - infinitely thin sheet; Case II quasi-uniform plasma sheet with no internal magnetic field; Case III – current sheet where the in-plane field gradually transitions to plasma moving inward; Case IV - current sheet where the in-plane field gradually transitions to guide field. Small field divergence and small horizontal gradients are expected but not shown.

We conclude that equilibrium current sheets in 2D and 2.5D cannot have both finite length and finite thickness. Force balance cannot be achieved both inside and outside the sheet while at the same time satisfying force balance across the sheet boundary. The sheets must be singular if they have finite length.

## 3D CURRENT SHEETS WITH LINE-TYING

The situation is entirely different if a guide field is present and variations are allowed in the out-of-plane direction. In this fully 3D case, the guide field can become curved, which introduces a tension force that was not previously present. Suppose the guide field is line-tied at two ends, i.e., the end points are held at fixed positions. As plasma flows away from the middle of the sheet toward the Y-points, the frozen-in guide field bows outward, as shown schematically in Figure 5. This produces a tension force that opposes the flow. If the force is strong enough, it may balance the horizontal pressure gradient driving the flow and allow an equilibrium to be established, thus preventing a full collapse.

Consider the version of Case IV that includes the additional uniform guide field, $B_{g0}$. We take $B_{g0}$ to be substantially larger than the shear field component external to the sheet, $B_{x0}$ at point B in Figure 3, as appropriate for the coronal heating problem. The in-plane field vanishes at Y-points, so the pressure at point E is $B_{g0}^2/(8\pi)$. Because of the minimal in-plane tension between B and D, the pressure at D is approximately $(B_{g0}^2+B_{x0}^2)/(8\pi)$ and takes the form of an enhanced guide field. The horizontal pressure gradient along the center of the sheet is therefore approximated by $[B_{x0}^2/(8\pi)]/[\lambda/2] = B_{x0}^2/(4\pi\lambda)$, where $\lambda$ is the sheet length.

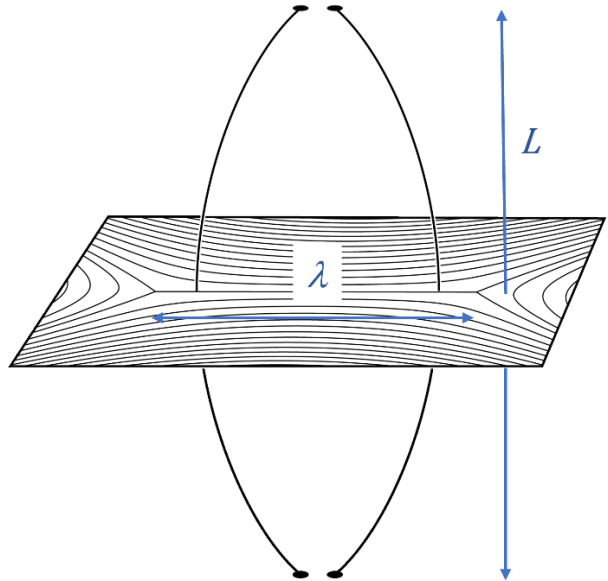

**FIGURE 5**  Schematic representation of a 3D current sheet that contains a guide field that is line-tied above and below. Horizontal flows from the middle of the sheet toward the Y-points have caused the field lines to become bowed. The length of the current sheet, $\lambda$, and separation of the line-tied footpoints, $L$, are indicated.

The horizontal tension force provided by the guide field is $B_{g0}^2/(4\pi R_c)$, where $R_c$ is its radius of curvature. The ratio of the tension to pressure forces is therefore $F_t/F_p \approx (\lambda/R_c)(B_{g0}/B_{x0})^2$. Equilibrium is possible as long as $R_c$ and/or $\lambda$ can adjust to make this ratio unity for a given magnetic shear ($B_{x0}/B_{g0}$).

Consider an equilibrium sheet in a field that is subjected to slow boundary driving. As the shear component of the field $B_{x0}$ increases, so too does the pressure gradient along the sheet. The plasma responds by slowly moving outward toward the Y-points, further bowing the guide field, increasing its tension, and maintaining a quasi-static force balance.[3]

There is an upper limit to the tension force in this scenario that occurs when an initially straight field line passing through the middle of the sheet ($x = 0$) is displaced all the way to a Y-point at the end. The two field lines in Figure 5 are not far from this limiting state. We can estimate the maximum tension force by assuming a circular arc for the guide field and taking $\lambda << L$, where $L$ is the separation of the line-tied ends (loop length in the coronal heating context). From simple geometry, the minimum radius of curvature is $R_{c,min} \approx (1/\lambda)(L/2)^2$. Substituting into the expression above for the tension-to-pressure force ratio, we obtain

$$\frac{F_t}{F_p} \leq 4\left(\frac{\lambda}{L}\frac{B_{g0}}{B_{x0}}\right)^2 . \tag{4}$$

Equilibrium is not possible when the right-hand side is less than unity because then tension is too weak to balance the pressure gradient. Systems with large footpoint separation (long loops) and strongly sheared fields are more prone to nonequilibrium, i.e., less able to support current sheets of finite thickness.

We can relate Equation 4 to a critical shear for a given sheet length and footpoint separation,

$$\frac{B_{x0}}{B_{g0}} > \left(\frac{B_{x0}}{B_{g0}}\right)_{crit} = \frac{2\lambda}{L}, \tag{5}$$

or to a critical footpoint separation for a given length and shear,

$$L > L_{crit} = 2\lambda\left(\frac{B_{g0}}{B_{x0}}\right). \tag{6}$$

When current sheets are free to lengthen – increase $\lambda$ – they will tend to do so to keep the conditions subcritical and maintain equilibrium. However, geometric constraints may sometimes prevent lengthening beyond a certain point. For example, the current sheets that separate elemental magnetic flux tubes that fill the corona[4] can be no longer than a tube "diameter" (the cross sections need not be circular, so "diameter" should not be taken literally). As chaotic photospheric flows twist and tangle the tubes, the magnetic shear across a mutual boundary increases and the sheet may transition from subcritical to supercritical conditions. It will then lose equilibrium, spontaneously collapse, and trigger magnetic reconnection. For a typical coronal loop length of 50,000 km and diameter of 2,000 km, the critical shear is 0.08, corresponding to a field rotation across the sheet of 10º. This is comparable to the roughly 20º needed to explain the energy budget of active regions (Klimchuk 2015). We note, however, that the elemental tubes that comprise a loop have smaller diameters, implying a smaller critical shear.

---

[3] Current layers in the simulation by Aulanier, Pariat, & Demoulin (2006) continue to thin for some time after the boundary driving ceases, indicating a non-trivial deviation from static conditions.

[4] The magnetic field of the high-$\beta$ photosphere is observed to be very clumpy, with a large fraction of the flux concentrated in small tubes of kilogauss strength. These tubes expand rapidly with height to become space filling in the low-$\beta$ corona above. There are at least 100,000 of these elemental tubes in a typical active region (Klimchuk 2015).

We caution that Equations 4, 5, and 6 are highly approximate and depend on the assumption that the guide field takes the shape of a circular arc. The numerical coefficients should be treated with special caution. Nonetheless, the dependencies on current sheet length, shear, and line-tied footpoint separation are intuitively very plausible.

## DISCUSSION

Using a simple force balance analysis, we showed why equilibrium cannot be achieved in 2D and 2.5D current sheets that have both finite length and finite thickness. We then estimated the conditions under which the tension of a line-tied guide field can facilitate 3D equilibrium in current sheets that are finite in all dimensions. We suggested that continuous 3D fields subjected to continuous driving at a line-tied boundary – the "Parker problem" – will contain current sheets of finite thickness until they reach a critical shear, whereupon they lose equilibrium, spontaneously collapse, and reconnect. The value of the critical shear is generally consistent with the observed heating requirements of solar active regions.

Equation 6 expresses the critical conditions in terms of the footpoint separation of the line-tied guide field, or loop length. A critical separation was also reported by Zhou et al. (2018). They investigated the so-called Hahm, Kulsrud, Taylor (HKT) problem involving the current sheet of a simply sheared force-free field that is locally squeezed and allowed to relax to a new equilibrium. The sheet is not bounded by Y-points, but the system is periodic in what would be the $x$ direction of our figures. We associate the $x$ dimension with a sheet length $\lambda$ because the essential physical effects included in our analysis are present in that direction in their simulations as well: (1) the in-plane field expands toward the periodic boundaries, thus providing an outward magnetic pressure gradient along the sheet, and (2) symmetric flows responding to the gradient will not cross the boundaries, thus limiting the bowing of the guide field. Note that Zhou et al. refer to the footpoint separation as a length, not to be confused with our current sheet length $\lambda$.

Zhou et al. found that the relaxed sheet is singular in 2.5D versions of their model and has finite thickness in 3D line-tied versions. The thickness decreases as the footpoint separation increases. There is a strong suggestion, but not definitive proof, that the 3D sheet becomes singular when the separation exceeds a critical size of $29\lambda$. In comparison, our Equation 6 predicts a critical separation of $11\lambda$. We do not consider this difference significant given the approximate nature of our derivation. Furthermore, the magnetic pressure gradient along the Zhou et al. sheet is smaller than it would be if the sheet terminated at Y-points. Less tension is therefore required for force balance, so we would expect a larger critical separation. On the other hand, the effects of line tying tend to be more pronounced near the line-tying boundaries, and the field line curvature will be relatively reduced away from the boundaries (Zweibel & Boozer 1985; Robertson, Hood, & Lothian 1992). This violates our assumption of a circular shape for the guide field and implies a reduced critical separation or critical shear.

Other studies of 3D line-tied current sheets also find that the sheet thickness decreases with increasing separation of the footpoints (Longcope & Strauss, 1994; Baty 1997; Huang, Bhattacharjee, & Zweibel 2010; Craig & Pontin 2014). To our knowledge, only we and Zhou et al. (2018) have proposed a critical separation beyond which the thickness plummets to zero.

It should be noted that our simple analysis assumes a planar current sheet. The tension that we describe arises from a curved guide field within the plane, and it exerts a force in the x-direction. If the current sheet were itself bowed, there would be an additional guide field curvature force in the y-direction that we do not consider. This can be the case, for example, when twisted flux tubes become kink unstable. Baty (1997) suggests that this different curvature may help prevent singularities from forming. Whether it would extend the critical footpoint separation (loop length) to larger values – or even eliminate it altogether – has yet to be determined. We are skeptical, however, because the force is directed perpendicular to the sheet and so is unable to balance the sheet-aligned pressure gradients that are the root of the problem.

The existence of a critical shear for loss of equilibrium and current sheet collapse is the most important outcome of our work. It offers a possible new explanation for reconnection onset, which is crucial for explaining a wide range of phenomena, including but not limited to coronal heating. It remains to be determined whether current sheets survive as they thin toward the critical value. The tearing instability can be very fast even in relatively thick sheets under coronal conditions (Pucci & Velli 2014; Leake, Daldorff, & Klimchuk 2020). Fast reconnection may set in before the critical shear is reached. We consider Equation 5 to be highly promising, but it must be rigorously and quantitatively evaluated. High resolution numerical simulations will be especially important.

## ACKNOWLEDGEMENTS

This work was supported by the GSFC Heliophysics Internal Scientist Funding Model competitive work package program and by a grant from the NASA Heliophysics Living With a Star Science Program. We are grateful to Yi-Min Huang and Spiro Antiochos for helpful discussions and to the referees for comments that led to a significantly modified and improved paper.

# REFERENCES

Antiochos, S. K. (1987). The Topology of Force-Free Magnetic Fields and its Implications for Coronal Activity. *ApJ* 321, 886-894.

Aulanier, G., Pariat, E., and Demoulin, P. (2005). Current Sheet Formation in Quasi-Separatrix Layers and Hyperbolic Flux Tubes. *A&A* 444, 961-976

Aulanier, G., Pariat, E., Demoulin, P., and DeVore, C. R. (2006). Slip-Running Reconnection in Quasi-Separatrix Layers. *Solar Phys*. 238, 347-376.

Baty, H. (1997). The Form of Ideal Current Layers and Kink Instability in Line-tied Coronal Loops. *A&A*. 318, 621-630.

Chen, P. F. (2011). Coronal Mass Ejections: Models and Their Observational Basis. *Living Rev. Solar Phys*. 8, 1.

Craig, I. J. D. and Pontin, D. I. (2014). Current Singularities in Line-Tied Three-Dimensional Magnetic Fields. ApJ 788, 177-185. doi: 10.1088/0004-637X/788/2/177

Craig, I. J. D. and Sneyd, A. D. (2005). The Parker Problem and the Theory of Coronal Heating. *Solar Phys*. 232, 41-62.

Green, R. M. (1965). IAU Symp. 22, 398.

Harris, E. G. (1962). On a Plasma Sheath Separating Regions of Oppositely Directed Magnetic Field. *Nuovo Cim* 23, 115. doi: 10.1007/BF02733547

Huang, Y.-M., Bhattacharjee, A., and Zweibel, E. G. (2010). Effects of Line-Tying on Magnetohydrodynamic Instabilities and Current Sheet Formation. *Phys. Plasmas* 17, 055707. doi: 10.1063/1.3398486

Huang, Y.-M., Hudson, S. R., Loizu, J., Zhou, Y., and Bhattacharjee, A. (2022). Numerical Study of Delta-Function Current Sheets Arising from Resonant Magnetic Perturbations. *Phys. Plasmas* 29, 032513. doi: 10.1063/5.0067898

Kazachenko, M. D., Albelo-Corchado, M. F., Tamburri, C. A., and Welsch, B. T. (2022). Short-term Variability with the Observations from the Helioseismic and Magnetic Imager (HMI) Onboard the Solar Dynamics Observatory (SDO): Insights into Flare Magnetism. *Solar Phys*. 297, 59. doi: 10.1007/s11207-022-01987-6

Klimchuk, J. A. (2015). Key Aspects of Coronal Heating. *Phil. Trans. R. Soc. A* 373, 201400256. doi: 10.1098/rsta.2014.0256

Leake, J. E., Daldorff, L. K. S., and Klimchuk, J. A. (2020). The Onset of 3D Magnetic Reconnection and Heating in the Solar Corona. *ApJ* 891, 62. doi: 10.3847/1538-4357/ab7193

Longcope, D. W. and Strauss, H. R. (1994). The Form of Ideal Current Layers in Line-Tied Magnetic Fields. *ApJ* 437, 851-859.

Low, B. C. (1992). Formation of Electric-Current Sheets in the Magnetostatic Atmosphere. *A&A* 253, 311-317.

Mikic, Z., Schnack, D. D., and Van Hoven, G. (1989). Creation of Current Filaments in the Solar Corona. *ApJ* 338, 1148-1157.

Ng, C. S. and Bhattacharjee, A. (1998). Nonequilibrium and Current Sheet Formation in Line-Tied Magnetic Fields. *Phys Plasmas* 5, 4028-4040. Doi: 10.1063/1.873125

Parker, E. N. (1972). Topological Dissipation and the Small-Scale Fields in Turbulent Gases. *ApJ* 174, 499-510.

Pontin, D. I. and Hornig, G. (2015). The Structure of Current Layers and Degree of Field-Line Braiding in Coronal Loops. *ApJ* 805, 47-59. doi: 10.1088/0004-637X/805/1/47

Priest, E. R. (1985). The Magnetohydrodynamics of Current Sheets. *Rep. Prog. Phys*. 48, 955.

Pucci, F., and Velli, M. (2014). Reconnection of Quasi-Singular Current Sheets: The "Ideal" Tearing Mode. *ApJ Lett* 780, L19. doi: 10.1088/2041-8205/780/2/L19

Raouafi, N. E., Patsourakos, S., Pariat, E., Young, P. R., Sterling, A. C. et al. (2016). Solar Coronal Jets: Observations, Theory, and Modeling. *Space Sci. Rev*. 201, 1. doi: 10.1007/s11214-016-0260-5

Robertson, J. A., Hood, A. W., & Lothian, R. M. (1992). The Evolution of Twisted Coronal Loops. *Solar Phys*. 137, 273-292

Syrovatsky, S. I. (1971). *Sov. Phys. JETP* 33, 933.

van Ballegooijen, A. A. (1985). Electric Currents in the Solar Corona and the Existence of Magnetostatic Equilibrium. *ApJ* 298, 421-430.

Wilmot-Smith, A. L., Hornig, G., and Pontin, D. I. (2009). Magnetic Braiding and Parallel Electric Fields. *ApJ* 696, 1339-1347. doi: 10.1088/0004-637X/696/2/1/1339

Zhou, Y., Huang, Y.-M., Qin, H., and Bhattacharjee, A. (2018). Constructing Current Singularity in a 3D Line-tied Plasma. *ApJ* 852, 3-11. doi: 10.3847/1538-4357/aa9b84

Zweibel, E. G., & Boozer, A. H. (1985). Evolution of Twisted Magnetic Fields. *ApJ*, 295, 642-647.

Zweibel, E. G. and Li, H.-S. (1987). The Formation of Current Sheets in the Solar Corona. *ApJ* 312, 423-430.